\documentclass[twocolumn]{aastex62}
\usepackage{multirow}

\graphicspath{{./}{figures/}}
\submitjournal{ApJ}
\shorttitle{Flaring Li-rich giant}
\shortauthors{Singh et al. 2024}
\begin{document}
\title{Study of a red clump giant, KIC~11087027, with high rotation and strong infrared excess - Evidence of tidal interaction for high lithium abundance}

\correspondingauthor{Raghubar Singh, Gang Zhao}
\email{raghubar@bao.ac.cn,gzhao@nao.cas.cn}
\author[0000-0001-8360-9281]{Raghubar Singh}
\affiliation{National Astronomical Observatories, Chinese Academy of Sciences, 20A Datun Road, Chaoyang District, Beijing, China (100101)}

\author[0000-0002-4282-605X]{Anohita Mallick}
\affiliation{Indian Institute of Astrophysics, 560034, 100ft road Koramangala, Bangalore, India}
\affiliation{Pondicherry University, R. V. Nagara, Kala Pet, 605014, Puducherry, India}

\author[0000-0001-9246-9743]{Bacham E. Reddy}
\affiliation{Indian Institute of Astrophysics, 560034, 100ft road Koramangala, Bangalore, India}

\author[0000-0002-4331-1867]{Jeewan C. Pandey}
\affiliation{Aryabhatta Research Institute of Observational Sciences (ARIES), Manora Peak, Nainital 263001, India}

\author{Gang Zhao}
\affiliation{National Astronomical Observatories, Chinese Academy of Sciences, 20A Datun Road, Chaoyang District, Beijing, China (100101)}

\begin{abstract}
This paper presents results from Kepler photometric light curves and high-resolution spectroscopic study of a super Li-rich giant KIC11087027. Using the light curve analysis, we measured the star's rotational period P$_{\rm rot}$=30.4$\pm$0.1~days, which translates to rotational velocity V$_{\rm rot}$=19.5 $\pm$ 1.7~km s$^{-1}$. Star's location in the HR-diagram, derived values of $^{12}C/^{13}C$ = 7$\pm$1 and $[C/N]=-0.95\pm 0.2$, and the inferred asteroseismic parameters from secondary calibration based on spectra suggest  star is a low-mass red clump giant in the He-core burning phase. Using Gaia data, we found evidence of variation in radial velocity and proper motion, indicative of presence of an unresolved binary. The large V$_{\rm rot}$ is probably a result of tidal synchronization combined with the after-effects of He-flash, in which the size of the star is reduced significantly. The simultaneous presence of features like high rotation, very high Li abundance, strong dust shell, and strong flares in a single star is relatively uncommon, suggesting that the star experiencing tidal synchronization has recently undergone He-flash. The results pose a question whether the binary interaction, hence the high rotation, is a prerequisite for dredging-up of the high amounts of Li from the interior to the photosphere during or immediately after the He-flash event.     

\end{abstract}

\keywords{low mass stars --- rotation --- stellar flares --- Li-rich giants---mass loss---stellar activity}

\section{Introduction}
A small group of red giants show very high Li abundances contrary to the general understanding that the Li gets destroyed in stars. This has been an anomaly ever since the discovery of the Li-rich giant by \cite{wallerstein1982}. Recent studies demonstrated that high lithium abundance among red clump (RC) giants is common \citep{Kumar2020} and showed that the He-flash event, immediately preceding the RC, holds the key for Li enhancement \citep{Kumar2011a, Casey2019, Deepak2019,singh2019,Yan2021}. The average Li abundance of RC giants, post He-flash, is about a factor of 40 more than their counterparts on the upper red giant branch (RGB), closer to the RGB tip \citep{Kumar2020}. In addition, the study by \cite{Singh2021} showed that A(Li) evolves with the g-mode period spacing ($\Delta \Pi_{1}$) evolution. During transition phase from the degenerate core to the convective core He burning, the ($\Delta \Pi_{1}$) value increases. As a result, very high A(Li) values are mostly seen at relatively lower ($\Delta \Pi_{1}$) values, i.e., immediately after the He-flash. The normal Li RC giants are seen at high $\Delta \Pi_{1}$ values. From this, they hypothesised that the super Li-rich (SLR) giant  phase (A(Li)\footnote{A(Li)=12+$\rm log(\frac{N(Li)}{(N(H)})$}$>$3.2~dex) is a short-lived phenomenon, and giants with very high Li abundance have undergone Li enhancement very recently, suggesting the He-flash as the source of Li enhancement. Further, the study by \cite{Mallick2023} showed  occurrence of Li-rich giants only among low-mass (M$\leq $ 2M$_{\odot}$) giants and none among intermediate (M$>$2M$_{\odot}$) mass RC giants providing an indirect evidence that Li enhancement occurs during the He-flash as the He-flash expected to happen in only  low-mass giants \citep{MillerBertolami2020}. We also note that there are Li abundance studies among cluster giants. Some of the super Li-rich giants in the clusters found to be  on the RGB \citep{Sanna2020,Nagarajan2023, Tsantaki2023}. If these giants are indeed ascending the RGB first time then there may be multiple ways for Li-enrichment in giants.

However, it is not understood what physical mechanism is responsible for the production and transportation of Li to the photosphere. Recent attempts to explain the physical process include internal gravity waves generated during He-flash \citep[e.g.][]{Schwab2020, Jermyn2022} and thermohaline mixing \citep{GaoJ2022}. There are proposals in literature to drive the convection and mixing of A(Li) with the upper atmospheres, like the differential rotation due to faster core rotation \citep{Fekel1993, Simon1989}. Stars may get faster rotations due to tidal interaction in binary systems driving the desired mixing process\citep{Denissenkov2004, Casey2019}. Any external events like mergers or tidal interactions with companions may spin-up the star, generating a strong magnetic field and creating circumstellar envelopes. One would expect all or most of these features if the Li enhancement occurred recently due to external events like mergers. 

In this paper, we analysed high-resolution spectra of KIC11087027 for the first time. The star is a super Li-rich giant with features such as high rotation, infrared excess (IR-excess) and high chromospheric activity, indicating some recent binary interaction. 

\section{ Observations and Data Reduction}
KIC11087027 was found in the sample of 12500 stars common in LAMOST \citep{Zhao2006,Cui2012,Zhao2012} and Kepler \citep{Borucki2010} field studied by \cite{singh2019}. This star was excluded in \cite{singh2019} because of the absence of oscillation modes in Kepler Power Density Spectra (PDS). The
star has peculiar features like the presence of rotational modulation, flares, and IR-excess. To supplement the photometric data, we obtained high-resolution (R=60,000) optical spectra using the 2.0-m Hanle Chandra Telescope (HCT) equipped with Hanle Echelle Spectrograph (HESP)\footnote{\url{https://www.iiap.res.in/hesp/}} for abundance analysis. We took three frames, each with a 40-minute exposure. The spectra cover the wavelength range from 3700 to 9300~\AA. We obtained calibration image frames - bias, flat, and Th-Ar lamp for wavelength calibration. We also obtained high-rotation hot star spectra for removing telluric lines. We followed the standard data reduction procedure for Echelle spectral reduction using IRAF \footnote{\url{https://iraf-community.github.io/}}. 
The resultant spectral image has a signal-to-noise ratio (SNR) of 86. The extracted spectrum was wavelength calibrated and continuum fitted for further analysis and derivation of stellar atmospheric parameters and elemental abundances.        
\begin{figure*}
    \centering
    \includegraphics[width=\textwidth]{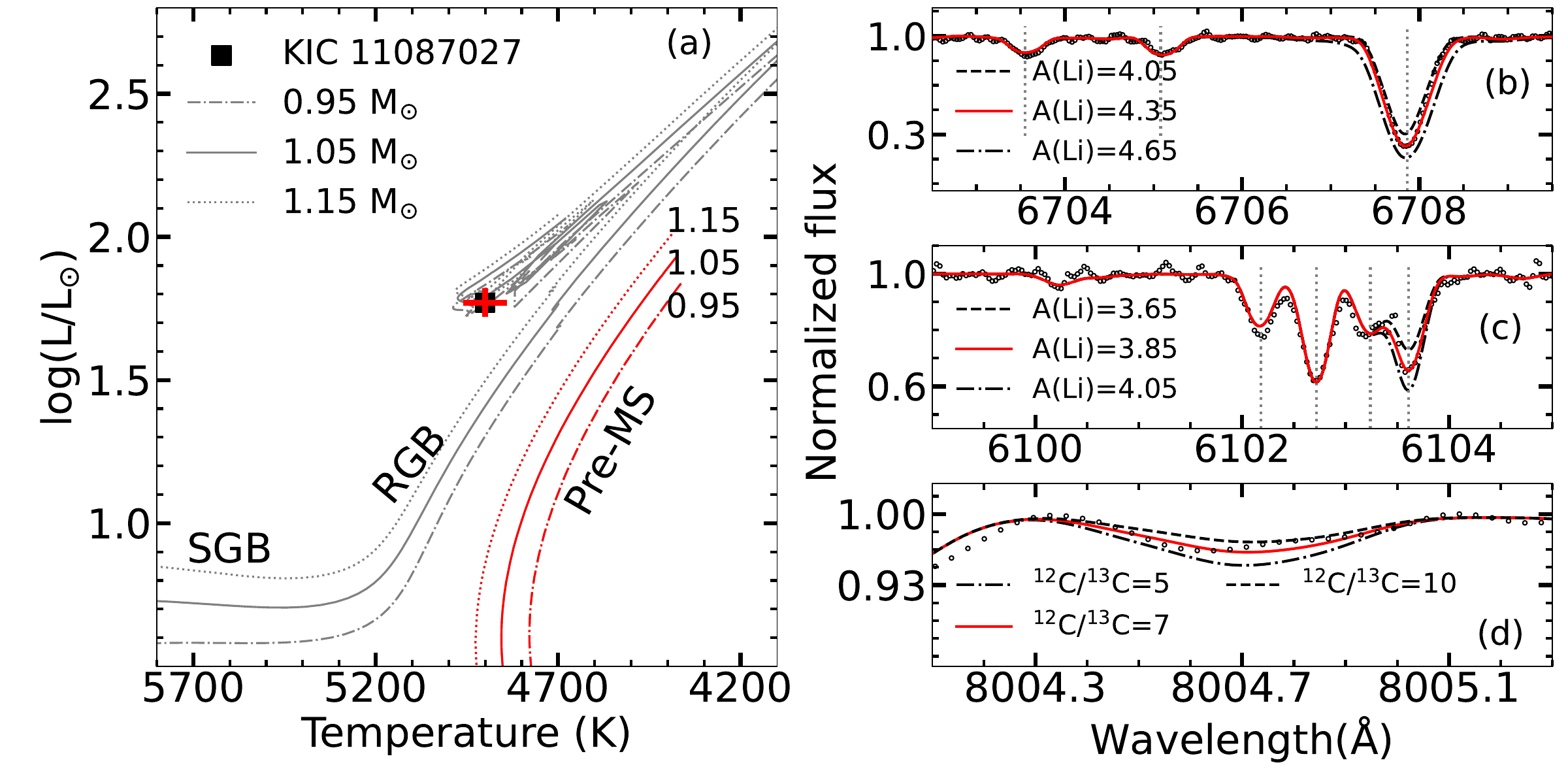}
    \caption{(a): Location of KIC11087027 (solid square), in the Hertzsprung-Russell diagram along with MESA-MIST \citep{Paxton2011, Dotter2016} evolutionary tracks for [Fe/H]=-0.55~dex with different masses. Panels
    (b),(c), and (d) are the spectral synthesis of Li resonance lines at 6707~\AA\, Li subordinate line at 6103~\AA\, and CN red line band near 8004~\AA, respectively.
    }
    \label{fig:hrd}
\end{figure*}

\section{Analysis and results}

\subsection{Stellar parameters and Abundances}
Stellar parameters (T$_{\rm eff}$, log~$g$, [Fe/H], Vsini, $\xi_{\rm t}$), see Table ~\ref{tab:tab1}, are derived from high-resolution spectra. We used FASMA\footnote{\url{https://github.com/MariaTsantaki/fasma-synthesis}} code \citep{Maria2018} for spectrum synthesis to derive stellar parameters. Star's luminosity is derived using the Gaia Gmag \citep{Gaia2016,Gaia2018} with appropriate bolometric correction \citep{Andrae2018} and the parallax given in Gaia. Extinction coefficient in G band, A$_{\rm G}$, is calculated by $dustapprox$ code  \citep{Fouesneau_dustapprox_2022} \footnote{\url{https://mfouesneau.github.io/dustapprox}}. Observed Heliocentric radial velocity is derived by cross-correlating the continuum-fitted spectra with the template spectra of Arcturus.

Elemental abundances and carbon isotopic ratio are derived by matching the predicted spectra with the observed ones. Model spectra are produced using the local thermodynamic equilibrium (LTE) stellar atmospheric models \citep{CastelliKuruxz2004} and the 1-D radiative transfer code MOOG \citep{Sneden1973}. The atomic and molecular line list is taken from Linemake\footnote{\url{https://github.com/vmplacco/linemake}} code \citep{Placco2021}. LTE Li abundance measured from the Li resonance line at 6707.78~\AA\, is A(Li)=4.35$\pm$0.10 dex and from the subordinate line at 6103~\AA\, is A(Li)=3.85$\pm$0.1~dex. The Li abundance is corrected for NLTE effects by following correction values provided by \cite{LindK2009} (Table~\ref{tab:tab1}). Carbon abundance is derived from atomic Carbon lines at 5086 and 5384.3~\AA\, whereas the  Nitrogen abundance is derived from $\rm ^{12}C^{14}N$ molecular line at 6486.4~\AA. The carbon isotopic ratio is measured from $^{13}$C$^{14}$N molecular line in the red band of the spectra at 8004.72~\AA. Errors in Li abundance and carbon isotopic ratio are derived as the quadratic sum of uncertainties in stellar parameters (T$_{\rm eff}$, log~$g$, [Fe/H], $\xi_{\rm t}$) and SNR. Abundances of other elements are derived using equivalent widths.

\begin{table}
\caption{Stellar parameters of KIC11087027. Elemental abundance w.r.t. to the Sun from \cite{Asplund2009}. }
\label{tab:tab1}
\centering
\begin{tabular}{lll}
\hline
Name & KIC11087027 & reference \\
\hline
R.A. & 19:33:14.212 & \\ 
DEC. & +48:41:50.83 & \\
V (mag) & $ 11.23 \pm 0.09$ & APASS  \\
A$_{\rm G}$ (mag)& $0.11$ & this work \\
Parallax (mas) & $0.86 \pm 0.05$ & Gaia DR3\\
RUWE& $4.38$& Gaia DR3\\
P$_{\rm rot}$ (Days)& $ 30.38 \pm 0.12 $ & this work \\
V$_{\rm rot}$(km s$^{-1}$) & 19.5 & this work \\
Vsini(km s$^{-1}$) & $ 10.1 \pm 0.2$& this work \\
Inclination angle & 35$^{\circ}$ & this work \\
Number of flares & 15 & this wok\\
T$_{\rm eff}$ [K] & $4920 \pm 60$ & this work\\
log~$g$ & $ 2.64 \pm 0.05 $ & this work\\
$\rm [Fe/H]$ & $ -0.54 \pm 0.05 $ & this work\\
$\xi_{t}$(km s$^{-1}$)& $0.95 \pm 0.05$ & this work\\
$\log(L/L_{\odot})$ & $ 1.77 \pm 0.05$ & this work\\
Mass (M$ _{\odot}$) & $ 1.014 \pm 0.2$& this work\\
Radius (R$ _{\odot}$) & $11.69 \pm 1$ & this work\\
RV$ _{\rm Gaia}$ (km s$^{-1}$) & $26.70 \pm 0.38 $ & Gaia DR3\\
RV$ _{\rm LAMOST}$ (km s$^{-1}$) & $ 22.71 \pm 4.11 $ & Luo+2018\\
RV$ _{\rm HESP, Helio}$ (km s$^{-1}$) & $19.98 \pm 0.26 $ & this work\\
A(Li)$_{\rm NLTE}$(6103~\AA) & $ 3.95 \pm 0.1$ & this work\\
A(Li)$_{\rm NLTE}$(6707~\AA) & $ 3.85 \pm 0.1$ & this work\\
$\rm [CI/Fe]$ & $ -0.18 \pm 0.1$ & this work\\
$\rm [NI/Fe]$ & $ 0.77 \pm 0.2$ & this work \\
$\rm [NaI/Fe]$ & $ 0.01 \pm 0.15$ & this work\\
$\rm [MgI/Fe]$ & $ 0.03 \pm 0.05$ & this work\\
$\rm [AlI/Fe]$ & $ 0.01 \pm 0.25$ & this work\\
$\rm [SiI/Fe]$ & $ -0.02 \pm 0.3$ & this work\\
$\rm [CaI/Fe]$ & $ -0.03 \pm 0.3$ & this work\\
$\rm [ScII/Fe]$ & $ -0.09 \pm 0.1$ & this work\\
$\rm [TiII/Fe]$ & $ 0.0 \pm 0.1$ & this work\\
$\rm [VI/Fe]$ & $ 0.13 \pm 0.22$ & this work\\
$\rm [CrI/Fe]$ & $ -0.02 \pm 0.3$ & this work\\
$\rm [CoI/Fe]$ & $ -0.28 \pm 0.18$ & this work\\
$\rm [NiI/Fe]$ & $ -0.34 \pm 0.21$ & this work\\
$\rm [ZnI/Fe]$ & $ 0.23 \pm 0.2$ & this work\\
$\rm [SrI/Fe]$ & $ -0.51 \pm 0.2$ & this work\\
$\rm [BaII/Fe]$ & $0.04 \pm 0.06$ & this work\\
$\rm [NdII/Fe]$ & $ 0.22 \pm 0.1$ & this work\\
$\rm [EuII/Fe]$ & $ 0.15 \pm 0.08$ & this work\\
$\rm ^{12}C/^{13}C $ & $7 \pm 1$& this work\\
\hline
\end{tabular}
\end{table}

\subsection {Evolutionary Phase}
The evolutionary phase of a star is determined using the star's location in the HR diagram combined with evolutionary tracks,  and using abundances ratios [C/N] and $\rm ^{12}C/^{13}C$. As shown in Fig~\ref{fig:hrd}, the star's luminosity and T$_{\rm eff}$ places it in a region in the HR diagram that overlaps with the 
red clump region. Similarly, observed [C/N]= -0.95$\pm$0.22 and [Fe/H]= -0.55$\pm$0.05, locates KIC11087027 in the red clump region in [C/N] vs [Fe/H] diagram \citep[see, Fig~5 top panel in][]{Hawkins2018}. The very low value of $\rm ^{12}C/^{13}C=7\pm1$ suggests that the star is highly evolved and, at least, beyond the luminosity bump. Though we have Kepler photometric data for 18 quarters, the power density spectrum (PDS) of KIC11087027 shown in Figure~\ref{fig:flarerot}(b) do not show Gaussian power excess with oscillation modes. Inhibition of oscillation modes is caused by enhanced magnetic activity \citep{Chaplin2011, Gaulme2014}. As a result, we resorted to adopt asteroseismic parameters using secondary calibrations by \citep{Wang2023}
who have derived  $\rm \Delta \Pi_{1}=187$~s and $ \rm \Delta \nu=4.4~\mu Hz$ using spectral indices from LAMOST spectra. The values suggest KIC11087027 is a red clump giant in He-core burning phase \citep{Bedding2011}.

\subsection{Detection of stellar flare and measuring flare energy}
\begin{figure*}
    \centering
    \includegraphics[width=\textwidth]{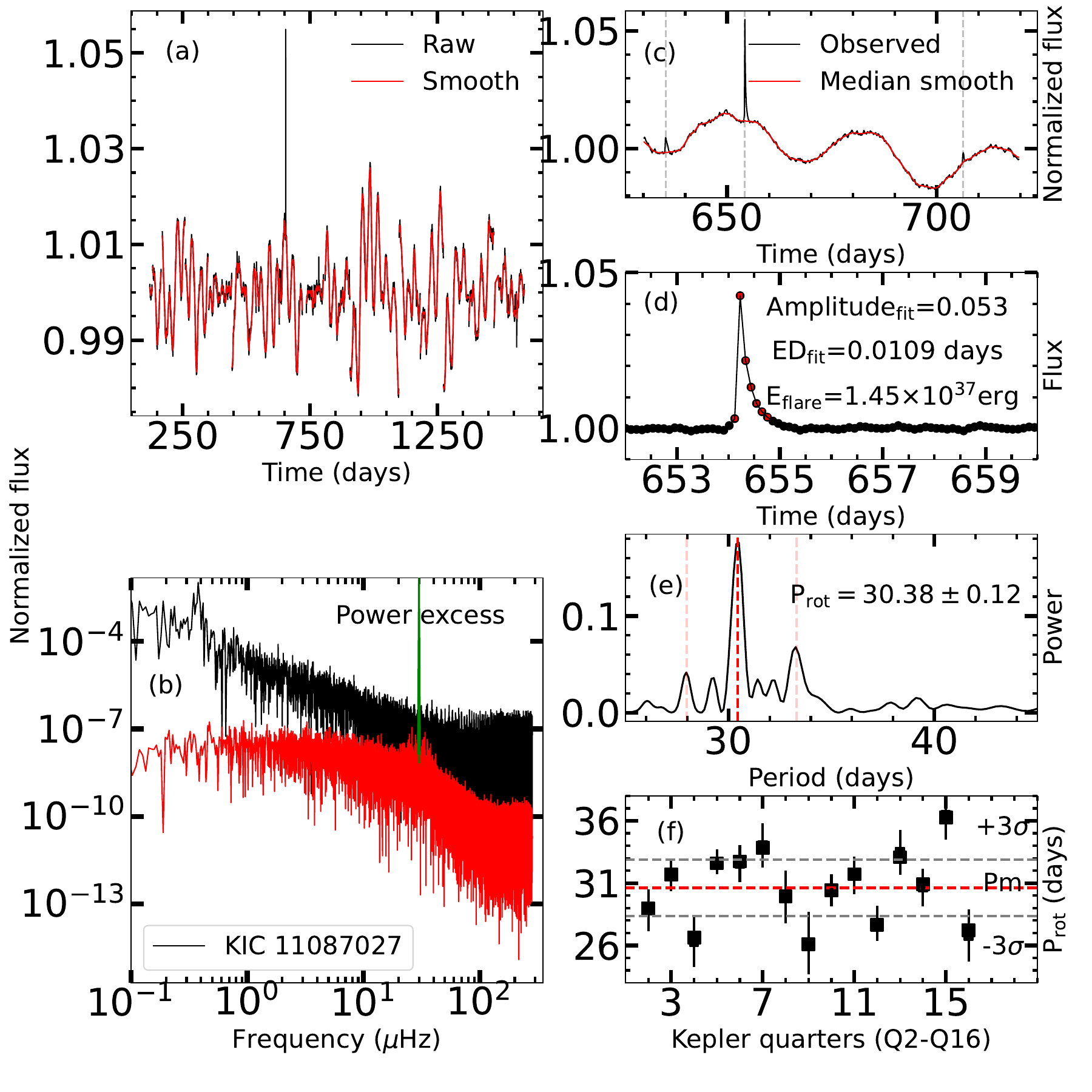}
    \caption{
    Kepler light curve analysis of KIC11087027: a)  black is raw and red is a median smoothed light curve. b) comparison of the power spectral density of KIC11087027 (black) with a similar mass star KIC12645107 (red) with no activity. c) 7th quarter Kepler LC of KIC11087027 and three flares are marked.  
    (d) The detrended light curve in the flare region and fitted flare model. Measured flare parameters are given in the figure. Panel (e): The Lomb-Scargle periodogram of the entire 18-quarter light curve. The vertical red dashed line is measured period. 
    (f): The rotation period measured in different quarters (Q2---Q16) gives different values. }
    \label{fig:flarerot}
\end{figure*}

Stellar flares are manifestations of magnetic energy. Flares are common features among main sequence stars but uncommon among red giants due to reduced magnetic field. Sometimes, stellar superflares may arise in binary companions or due to recent activity with stellar companion-like planets \citep{Cuntz2000}. We have investigated long cadence (29.4 minutes) white light photometric data from the Kepler mission obtained over four years. The visual inspection of the light curve showed many fast rises and exponential decay-like profiles in amplitude, suggesting the presence of flares. We used $\rm FLATW^{,}RM $ code \citep{Vida2018} to detect stellar flares and measure flare parameters including equivalent duration and amplitude. We discovered 15 white light flares (WLFs) in KIC11087027 of the duration of a few hours each; see Figure~\ref{fig:flarerot}~(b). 
 WLF energy is derived from the equivalent duration (ED) of flare and quiescent luminosity. Quiescent luminosity in the Kepler wavelength band is measured by integrating the convolution of bolometric flux ($F_{\lambda}$) and Kepler response function ($\rm S_{Kp}$). 
$$E_{flare}=4 \pi R^{2}\times ED\int_{\lambda_{1}}^{\lambda_{2}}F_{\lambda}*S_{Kp}d\lambda$$ 
Here $\lambda_{1}$ and $\lambda_{2}$ are the lower and upper wavelength limits of the Kepler filter. Bolometric flux $F_{\lambda}$ for stellar parameters is taken from VOSA \footnote{\url{http://svo2.cab.inta-csic.es}} and Kepler response function $\rm S_{Kp}$ from Kepler website \footnote{\url{https://keplergo.github.io/KeplerScienceWebsite/}}. Energy of flares is in the range of $10^{35}-10^{37}$~erg, which makes KIC11087027 as super flaring star. 

\subsection{IR-excess and chromospheric activity}
We looked for infrared excess to understand whether the star had undergone any mass loss due to external merger events, He-flash event or binary interactions. We collected observed fluxes from near UV to far IR wavebands. 
The observed spectral energy distribution (SED) of KIC11087027 is shown in Fig~\ref{fig:halphased}(b). The observed SED is compared with the model SED of the star's derived stellar atmospheric model taken from Kurucz grid of flux models and is constructed using VOSA\footnote{\url{http://svo2.cab.inta-csic.es/theory/vosa/}}. Star shows significant IR-excess in near IR and far IR wavelengths, indicating an extended circumstellar environment due to episodic mass loss. Star also shows a double peak H$_{\alpha}$ emission profile as shown in Figure~\ref{fig:halphased}(a), which is asymmetric and variable, indicating the presence of chromospheric activity and mass motion in stellar atmosphere \citep{Dupree1984}. 

\subsection{SED modelling}
The 1-D radiative code DUSTY \citep{Ivezic1999} combined with the Kurucz model fluxes is used to derive dust parameters for a given set of atmospheric parameters and the assumed dust grain distribution. We have assumed a spherically symmetric dust shell with an oxygen-rich environment of warm silicate particles \citep{Draine1984} and standard MRN grain size distribution from \cite{Mathis1977}. A model SED is computed and compared with the observed SED (normalized to K$_s$-band fluxes) by varying dust parameters: inner dust temperature (T$_{\rm in}$), optical depth ($\tau_{\nu}$); dust shell relative thickness y=$\rm \left(\frac{R_{out}}{R_{in}}\right)$. 

\begin{figure*}
    \centering
    \includegraphics[width=\textwidth]{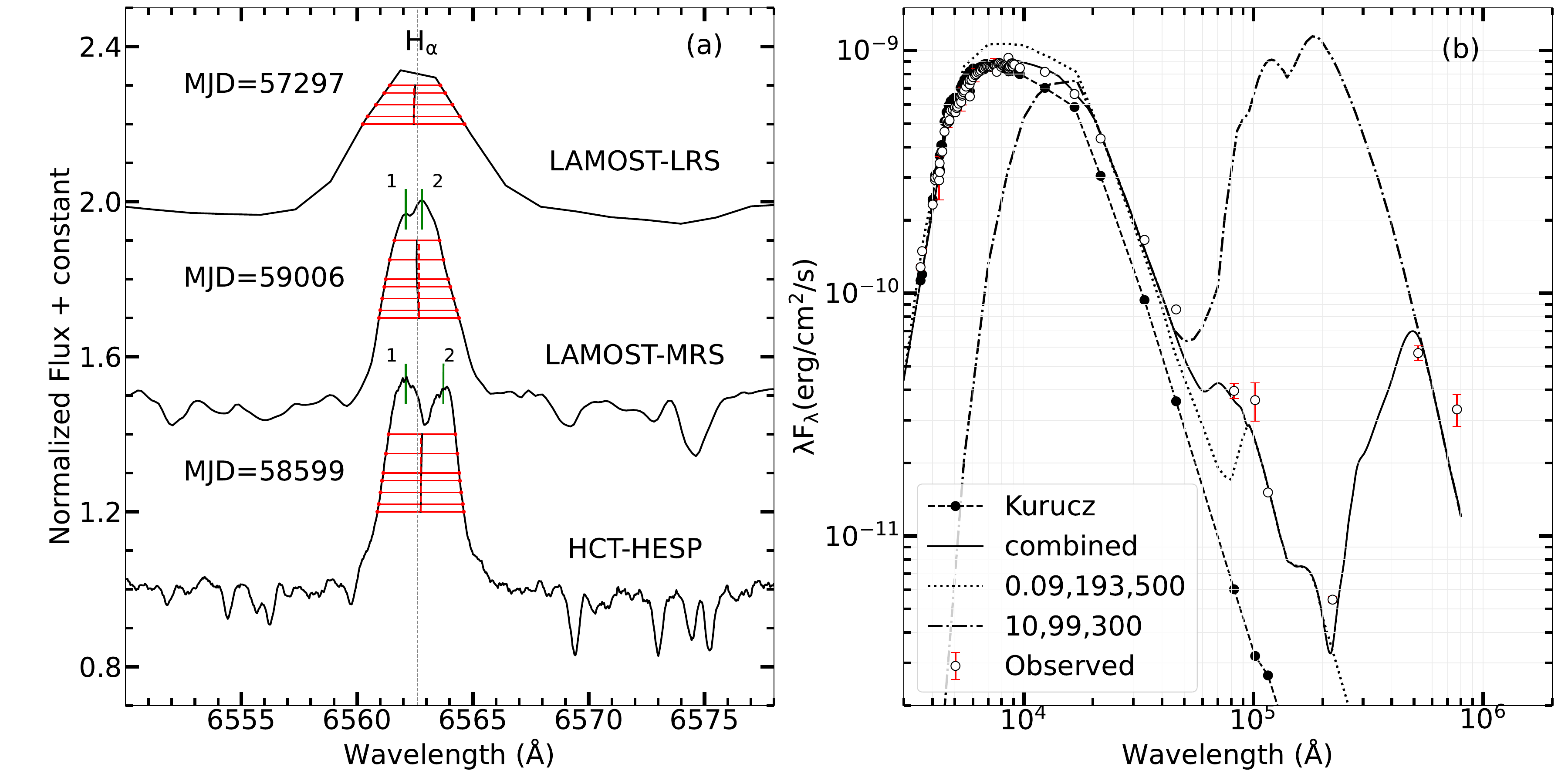}
    \caption{(a): H$_{\alpha}$ emission in KIC11087027. Three spectra are taken at different times and resolutions: LAMOST LRS (R=1800), MRS and HESP (R=7500, 60000). At higher-resolution, the two H$_{\alpha}$ emission peaks become visible.  
    (b): SED fitting to observed fluxes. 
    }
    \label{fig:halphased}    
\end{figure*}

We used a modified minimum chi-square test $\chi^{2}$-test for the best fit. $$\rm \chi^{2}=\sum \left(\frac{f_{obs}-f{mod}}{\sigma_{obs}}\right)^2/(N-p-1)$$ where  f$_{\rm obs}$, f$_{\rm mod}$ and $\sigma_{\rm obs}$ are observed flux, model flux and error in observed flux respectively. (N-p-1) represents the degrees of freedom, where N is the number of observed data points, and p is the number of independent parameters. Observed data is best fitted with two component dust shells: an inner hot and an outer cooler dust shell. The two-component dust shell best-fit model using DUSTY is shown in Fig.\ref{fig:halphased}(b). 

The dust shell inner radii and expansion velocities have been obtained using scaling relations. Other parameters of dust shell: mass loss rates ($\rm \dot M$), kinematic age ($\tau_{\rm d}$) and total mass (M$_{\rm d}$) are derived following \cite{Mallick2022} and are given in Table~\ref{tab:tabdusty}. The expected mass loss rate ($\dot M_{\rm R}$) is calculated using the modified \cite{Reimers1975} formula for derived stellar parameters (see Table~\ref{tab:tabdusty}).  

Stellar flares on a star's surface may lead to distinct inner warm and outer cool dust shells, coupled with episodic mass loss \citep{Osten2015}. Intense stellar radiation can lead to sublimation of dust grains closer to the star, creating an inner warm dust shell. The episodic events, shaped by flare variability may contribute to episodic mass loss from the two dust shells. 

\begin{table*}
\caption{Dust shell parameters of KIC11087027.}
\label{tab:tabdusty}
\centering
\begin{tabular}{cccccccccc}
\hline
&$\dot M_R$ (M$_{\odot}.yr^{-1}$)&T$_{\rm inner}$ & $\tau_{\nu}$ & T$_{\rm d}$ & V$_{\rm s}$ &$R_{\rm in}$ & t$_{\rm d}$ & $\dot M$ (M$_{\odot}\rm yr^{-1}$) & M$_{\rm d}$\\
&(Reimer's Law)& (K) &  & (K) & (km s$^{-1}$)& (cm) & (years) & (DUSTY) & (M$_{\odot}$)\\
\hline
\multirow{2}{*}{Inner Shell}&\multirow{4}{*}{6.68$\times$10$^{-11}$}&&&&&&&&\\
&  & 500 & 0.09 & 193 & 5.48 & 4.87$\times$10$^{13}$ & 2.82 & 5.78$\times$10$^{-9}$ & 8.26$\times$10$^{-8}$ \\
\cline{3-10}
\multirow{2}{*}{Outer Shell}&&&&&&&&&\\
&&300&10&99&1.41&2.45$\times$10$^{14}$&55.09&8.29$\times$ 10$^{-7}$& 2.32$\times$ 10$^{-4}$\\
\hline

\end{tabular}
\end{table*}

\subsection{Rotational velocity}
We estimated the star's rotational velocity, V$_{\rm rot}$, from spectra and photometric light curve data. The observed spectral width results from blending of other lines in the vicinity, instrumental broadening, and stellar phenomena like macroturbulence and rotation. We used two well-defined Fe~I lines at 6703 and 6705~\AA\ for extracting the rotation part of the line width by computing a set of profiles for given line abundance values, stellar parameters and instrumental broadening as measured from the Th-Ar calibration lines in the vicinity. The macroturbulence velocity, V$_{\rm mac}$=3.9~km s$^{-1}$, is calculated using calibration relation involving T$_{\rm eff}$ and log~g \citep{Hekker2007}. Using $\chi$$^{2}$-test we found the observed profiles best fit for Vsini=10.1~km s$^{-1}$. We also made use of Kepler photometric data to determine the period of the star. The data showed modulation in the light curve, probably due to large stellar spots that co-rotate with the stellar surface. We performed periodogram analysis by using the Lomb-Scargle method \citep{1982Scargle} in the Kepler data spanning over four years. Periods ranging from 26 to 33 days during different quarters, as shown in the bottom panel of Fig.\ref{fig:flarerot}(f), are found. The average period is found to be  30.38$\pm$0.12 days. Using the estimated star's radius, R=$11.69\pm1$~R$_{\odot}$ and considering this period as a rotational period (P$_{\rm rot}$), we found average $ V_{\rm rot}$=$ 19.5\pm1.7$~km~s$^{-1}$, which is quite large for an evolved star like KIC11087027. The quarter-to-quarter variation, with a median absolute deviation of 2.1, in the rotation is probably caused by latitudinal differential rotation \citep{Suto2022}.

\subsection{Binarity: Tidal synchronisation}
We did not find any resolved binary of KIC11087027 within 10$^{''}$ in the Gaia survey, and also, no signature of eclipse is present in the Kepler light curve. However, the presence of any unresolved companion can not be ruled out. The star has been observed at four different epochs in APOGEE, LAMOST, and HCT-HESP over a span of 7 years, and we found a small difference in the radial velocity (see Table~\ref{tab:tab1}). Gaia astrometry has been used for detecting unresolved companions of stars within 1$^{''}$, which can affect astrometric measurements. Gaia renormalized unit weight error (RUWE) value is greater than 1.4 for multiple system \citep{Ziegler2020} and RUWE of KIC11087027 is 4.38. This inflated value of RUWE indicates a close unresolved binary or multiple system, as a wide binary does not inflate RUWE. This is further supported by the proper motion anomaly (PMa) found for KIC11087027. Proper motion values of KIC11087027 measured by Tycho-2 in the updated UCAC4 catalog \citep{Zacharias2013} are $\mu_{\rm T}=(-2.1\pm0.7,-5.0\pm0.5)$ mas yr$^{-1}$ and Gaia DR3 are $\mu_{\rm G}=(-1.547\pm0.067,-7.289\pm0.059)$ mas yr$^{-1}$. Significant acceleration ($  \sigma=4.55$) is found in the $\delta$ direction with $ \mu_{\rm T}-\mu_{\rm G}=(-0.55\pm0.70, 2.29\pm0.50)$, a signature that the star may be an unresolved binary or has multiple systems. The absence of oscillation modes (see Figure \ref{fig:flarerot}b) and the presence of strong magnetic activity in low mass fast rotating evolved star \citep{Gaulme2014} suggests that the star KIC11087027 is probably in a tidally synchronised system. Other evidence is the star's high projected rotational velocity (Vsini=10~km s$^{-1}$) for its colour, (B-V)=1.04 \citep{demedeiros2002}. The observations suggest KIC11087027 is tidally locked with an unresolved close binary companion with an orbital period of 30.4~days.

\section{Discussion} 
Red giants with very high Li and strong IR-excess are rare because both properties seem transient and evolve differently. Though we do not have much information about IR-excess, the large surveys showed no evidence of IR-excess among giants \citep{Kumar2015}, and only a couple of Li-rich giants and SLRs are known to have IR-excess \citep{Mallick2022}. The general lack of IR excess among Li-rich giants indicates either no ejection of mass during the event that caused Li enrichment or the mass loss occurred, but the resultant IR-excess probably diluted faster than the Li depletion. In that sense, the star KIC11087027 is unusual, with very high Li and large IR-excess. This suggests Li-enrichment occurred recently, assuming the high Li and IR-excess result from a single event that is responsible for high Li among RC giants. Further, it is demonstrated that $\Delta \Pi_{1}$ is a proxy to the time evolution of a degenerate core into a convective core i.e., older RC giants with normal Li have relatively much higher $\Delta \Pi_{1}$ values \citep{Singh2021}. The estimated value of $\Delta \Pi_{1}$=187~s from spectra \cite{Wang2023} indicates that KIC~11087027 is a young RC and Li-enrichment occurred very recently. The star also shows very high stellar rotation (V$_{\rm rot}$=19.5~km s$^{-1}$), which causes the star to have high chromospheric activity with flares. 
What process might have created all the above signatures in a single star is unclear. If these properties are a consequence of a single event like He-flash and evolve at different time scales, one would expect stars with all these properties to be uncommon. Here, we also note, the study by \cite{Sneden2022} which shows a subtle correlation between the strength of chromospheric line at He~I 10830\AA\  and the high A(Li) in giants. They also show higher binary fraction among Li-rich giants compared to Li-poor implying some kind of binary interaction for Li production and chromospheric activity.

In the case of KIC 11087027, we discuss two plausible scenarios. One is an in-situ process in single-star evolution, and the other is the stars' binary evolution, where orbital synchronisation causes high rotation aiding high Li abundance. It is highly likely that He-flash holds the key to high Li anomaly among red clump giants \citep{Casey2019,Deepak2019,singh2019, Singh2021, Mallick2023}. Of course, we need to understand the nature of some of the very high Li-rich giants reported in clusters. These giants appear to be on the RGB and well before the RGB tip \citep{Sanna2020,Nagarajan2023} i.e before the He-flash event.  It is important to analyse the cluster  Li-rich giants using asteroseismology to find whether these are ascending the RGB. We will address this in the later study.

As per the RC Li-rich giants, the question remains about the physical process that mixes internally produced Li with the upper layers. Is He-flash the sole process? And does external aid, such as tidal locking, also play a role? Supporting the high levels of Li among red clump giants \cite{Schwab2020} constructed models in which the powerful gravity waves generated during the He-flash provide a conducive environment for large-scale convection. For this, they required diffusion coefficient D$_{\rm mix}$=10$^{11}$cm$^{2}$s$^{-1}$, which is a few orders more than expected in stars ascending the RGB. 

Let's assume Li gets produced inside the star through the Cameron-Fowler mechanism \citep{Cameron1971}, and the He-flash aids large-scale mixing. Though the models predict no observable feature like IR-excess due to the He-flash at the centre, it is not unreasonable to suggest mass loss occurs due to the flash, and the severity of mass loss depends on the intensity of the He-flash. The resultant IR-excess is diluted faster since it is a one-time event. This may be why red clump giants with IR-excess are rare. Importantly, all the SLRs or Li-rich giants don't need to have IR-excess as they may evolve at different time-scales. The red clump giants with IR-excess are found more likely to be super Li-rich or Li-rich \citep{Mallick2022}  suggesting that IR-excess dilutes faster than Li depletion. 

Concerning the high rotational velocity of KIC11087027, it is expected that the He-flash and the consequent reduction in star size by a factor of $\sim$10 will spin-up the star. Given the V$_{\rm rot}$ $\sim$0.7~km s$^{-1}$ for the giants at the RGB tip one would expect V$_{\rm rot}$ $\sim$ 7~km s$^{-1}$ for a young RC giant. This expected value due to a sudden drop in the size of the giant from the RGB tip to the RC is much smaller than the observed value of 19.5~km~s$^{-1}$. This implies high rotation in KIC11087027 and other Li-rich RC stars found in the literature probably arose due to external factors like mergers, or tidal locking. A primary giant at or near the RGB tip, which is tidally locked with its binary star, may have a larger rotational velocity than a single star. In the case of KIC11087027, observed V$_{\rm rot}$=19.5~km s$^{-1}$ may be the combination of tidal locking and the after-effect of the He-flash.

If we assume the star in tidal locking with close binary has slightly more V$_{\rm rot}$ say 2 km s$^{-1}$ rather the expected value of 0.7 km s$^{-1}$, the resulting post He-flash V$_{\rm rot}$ value matches with the observed high velocity. The large V$_{\rm rot}$ will increase the diffusion coefficient to 10$^{11}$ cm$^{2}$ s$^{-1}$ \citep{Casey2019} as D$_{\rm mix}\propto\Omega^{2}$. Models show that a significant increase in the value of D$_{\rm mix}$ may result in the observed lithium abundance in the photosphere\citep{Denissenkov2003}.

\section{Conclusion \label{conclusion}}
In this study, we reported results from Kepler photometric light curve and high-resolution spectral data. 
Based on variation in radial velocity and Gaia astrometric data, we suggest that KIC 11087027 has an unresolved companion. The observed high V$_{\rm rot}$=19.5~km s$^{-1}$ may be a result of a combination of tidal synchronization in a binary system and the effect of He-flash. The high rotation probably results in a high diffusion coefficient hence, the high Li abundance. The presence of several transient features like very high Li abundance, multiple dust shells, high rotation and strong flares provide evidence that either the star has undergone a merger-induced He-flash \citep{Mallick2022} or the tidally locked primary star has undergone the He-flash very recently. Also, to have a comprehensive understanding of origin of Li enhanced giants it is imperative to understand the nature of Li-rich giants in clusters which are reported to be ascending the RGB. 

\section{Acknowledgement}

We thank anonymous referees for useful comments which has improved our manuscript. This study is supported by the National Natural Science Foundation of China under grant No. 11988101, and the National Key R$\&$D Program of China No.2019YFA0405500. Guoshoujing Telescope (the Large Sky Area Multi-Object Fiber Spectroscopic Telescope LAMOST) is a National Major Scientific Project built by the Chinese Academy of Sciences. LAMOST is operated  and managed by the National Astronomical Observatories, Chinese Academy of Sciences. This work presents results from the European Space Agency (ESA) space mission Gaia. Gaia data are being processed by the Gaia Data Processing and Analysis Consortium (DPAC). Funding for the DPAC is provided by national institutions, in particular the institutions participating in the Gaia MultiLateral Agreement (MLA). The Gaia mission website is https://www.cosmos.esa.int/gaia. We gratefully acknowledge the entire team of Kepler space telescope which is funded by NASA’s Science Mission.

\bibliographystyle{aasjournal}

\end{document}